# A NoSQL Data-based Personalized Recommendation System for C2C e-Commerce

Tran Khanh Dang[1(✉)], An Khuong Vo[1(✉)] and Josef Küng[2]

[1] HCMC University of Technology, Vietnam National University, Ho Chi Minh City, Vietnam
`khanh@hcmut.edu.vn, voankhuong14@gmail.com`
[2] FAW Institute, Johannes Kepler University Linz, Linz, Austria
`jkueng@faw.jku.at`

**Abstract.** With the considerable development of customer-to-customer (C2C) e-commerce in the recent years, there is a big demand for an effective recommendation system that suggests suitable websites for users to sell their items with some specified needs. Nonetheless, e-commerce recommendation systems are mostly designed for business-to-customer (B2C) websites, where the systems offer the consumers the products that they might like to buy. Almost none of the related research works focus on choosing selling sites for target items. In this paper, we introduce an approach that recommends the selling websites based upon the item's description, category, and desired selling price. This approach employs NoSQL data-based machine learning techniques for building and training topic models and classification models. The trained models can then be used to rank the websites dynamically with respect to the user needs. The experimental results with real-world datasets from Vietnam C2C websites will demonstrate the effectiveness of our proposed method.

## 1 Introduction

With the increasing popularity of the Internet, e-commerce markets are developing rapidly these days. Among types of e-commerce, the B2C (Business-to-Customer) websites in which manufactures or retailers sell their products to consumers are occupying the largest market share. However, with the trend towards more economical and convenient transaction between consumers, many people are now shifting to using the C2C (Customer-to-Customer) websites. The C2C model brings lower costs and higher profits for buyers and sellers. Therefore, C2C websites are currently growing strongly in quantity as well as in quality. In Vietnam, as a proof for this trend, there are now hundreds of active C2C marketplaces with some big names such as *Cho Tot* [20], *Nhat Tao* [21], *Vat Gia* [22], etc. This poses problems known as information overload that it is difficult for users to choose the suitable websites to sell their items. However, most of the current decision support systems, specifically e-commerce recommendation systems, only directly benefit the buyers [1, 2].

In this paper, we are going to design a recommendation system which suggests suitable C2C websites to sell an item by applying machine learning techniques. The system is built in the context that the statistics, facts, and sale data is not provided. In this regard, the simplest way is selecting websites based on the category of the item. However, such category-based recommendation is not enough to model a specific item. After deeply studying the common features on some popular C2C websites, we focus on three features for processing: the description, the category, and the price of the item. Specifically, when a user provides the system those item's information, the system will recommend a ranked list of appropriate websites. In the construction of the C2C recommendation engine, one of the challenges we need to deal with is the unstructured text data on C2C websites, since the descriptions are usually written freely in user styles. To overcome this, we apply topic modeling [3, 4, 5], a form of text mining for identifying patterns in a corpus. We compute with topic modeling techniques to learn the latent subtopics in the descriptions to get the distribution of topics in each description and use the results as semantic features for further processing.

To personalize the recommendation system that can adapt to seller needs. We give two options for ranking websites. First, the websites are ranked in order of a specific criterion. For example, the websites are either ranked by the quantity or the average price of the similar items that are posted on each site. Second, the websites are ranked based on the whole input: the description, the category and desired selling price. For the first option, we measure the description similarity on the pre-processed semantic features and accumulate the results. For the second option, we build a classification model to predict a set of target websites for an item. The model must learn on single-label dataset but produce multiple labels. Moreover, the model must handle the data in which the same instance could belong to different labels since the same stuff could be posted to many similar websites. To solve this, we utilize the ensemble learning methods [6, 7], which have lots of learners that each learns a different aspect of the dataset and put them together to aggregate the results. The trained classifier, based on the voting scheme [8], ranks the websites in order of most voted.

The rest of the paper is organized as follows. We first introduce the related works in the Section 2. Section 3 goes into details of the proposed method. In section 4, we describe our experimental setup and present the results. Finally, Section 5 concludes the paper and provides some potential ideas for future work.

## 2    Related Works

There are a few research works about C2C e-commerce recommendation systems have been conducted. In [9], Guangyao proposed a recommendation model for C2C online trading. The method calculates the similarity between current active users and other users based on a preference matrix. The matrix is constructed by user's behaviors on C2C context in four criteria: browse, attention, bidding, and purchase. Then, a recommendation list is built for each user based on most important items for his neighbors. Also, three-dimensional collaborative filtering technique which is extended from traditional two-dimensional collaborative filtering is an approach for C2C recommendation system [10]. The method firstly calculates seller similarities using seller features, and fills the rating matrix based on sales relations and seller similarities. Next, it calculates the buyer similarities using historical ratings, and defines neighbors of each user to predict unknown ratings. Lastly, the system recommends the seller and product combinations with the highest prediction ratings to the target user. In [11], Bahabadi et al. proposed a solution based on social network analysis. The research focused on users and transactions, which formed the map and the network to incorporate link prediction techniques for building recommendation system using prior transactions of users, categories of items, rating of users and reputation of sellers.

To the best of our knowledge, this is the first attempt to construct C2C e-commerce recommendation system aiming at choosing appropriate websites for target items. Besides, this paper uses machine learning techniques such as topic modeling and ensemble learning, an approach that has not been systematically studied in previous research works about C2C e-commerce recommendation system.

## 3 Proposed Approach

In this section, we will describe in detail: (i) how the C2C e-commerce database is constructed for further data processing, (ii) how the unstructured text data, the item's description, is represented, (iii) how the websites are ranked based on specific criteria such as quantity or price using document similarity, (iv) how the websites are ranked based on the voting principle of the classifier.

### 3.1 The C2C e-Commerce Database

The local database of C2C e-commerce is constructed with the aim to provide an extensive and reliable representation of items that can be queried in a reasonable time. Our database for recommendation engine is illustrated in figure 1. First, the data is collected from many C2C e-commerce websites. Then, it is extracted and stored in a local database. Since there is a vast volume of data, local repository needs to perform queries in a short time. Besides, the database must allow flexibility in various document structures in many websites, as well as allow an easy adaption to integrate new dataset into the system, this database is in form of NoSQL database. We choose Elasticsearch [12], a distributed, open-source full-text search and analytical engine which can serve as a schema-less database. It provides horizontal scalability and automatically maintains fault tolerance and load distribution. The efficiency of Elasticsearch is due to its quick searching on inverted index [13] instead of searching the text directly. As Elasticsearch does not enforce document structure, single index can store documents with different fields. Based on this, we create each index for each website.

Once the data is in indexed, we run searches and aggregations to mine the data needed for learning algorithms in the recommendation engine. After the models are trained, they are used to compute the relevance of documents as well as to rank the websites for user's queries. Additionally, the trained models can help to organize, search, and classify web documents in the database more effectively.

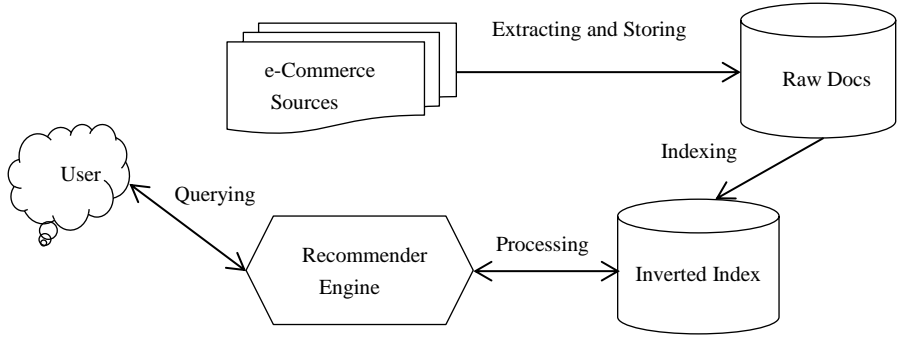

**Fig. 1.** The local C2C e-commerce database

### 3.2 Item's Description Representation

We first created a corpus using the descriptions of items from C2C e-commerce websites. First of all, each description is lowercased and tokenized, all the common words, as well as words that only appear once in all the descriptions are removed. Then descriptions are represented in bag-of-words (BOW) model [13]. Each document is represented by one vector where each vector element is represented by term frequency–inverse document frequency (tf-idf) weighting [13], denoted as:

$$tf * idf = f_{t,d} \log \frac{N}{n_t} \qquad (1)$$

where $f_{t,d}$ is the number of occurrences of term $t$ in document $d$, $N$ is the total number of documents in a corpus and $n_t$ is the number of documents where term $t$ appears. The tf-idf value increases proportionally to the number of times a word appears in the document, but is compensated by the frequency of the word in the corpus, which reduces the importance of terms that occur very frequently. However, this BOW model alone is not enough to depict the underlying semantic meanings or concepts of documents. For example, the phrases "Samsung S6 32gb like new" and "Galaxy S6 32gb 99%" are the same in meaning but barely identical vectors in BOW model. Also, the model generates very high dimensional features, over 10000 dimensions.

To improve efficiency, topic modeling [3, 4, 5] technique is proposed to fit the constructed tf-idf corpus. Topic modeling is an unsupervised approach, aiming at discovering the hidden thematic structure within text body. Using themes to explore the content of the description, we can reveal various aspects of it such as brand, status, hardware specification, etc. A topic is considered as a cluster of words that tend to contextually co-occur together. It connects words with similar meanings and distinguish the uses of words with multiple meanings. Besides, text representation with topic semantic space can greatly reduce feature dimension comparing to BOW feature. This paper compares two most common techniques, the probabilistic model, Latent Dirichlet Allocation (LDA) [14], and more recently, the non-probabilistic model, Non-Negative

Matrix Factorization (NMF) [15]. Both algorithms take as input the bag-of-words matrix $A \in R^{n \times m}$. Each algorithm attempts to produce two smaller matrices: a word to topic matrix $W \in R^{n \times k}$ and a document to topic matrix $H \in R^{k \times m}$, that when multiplied together reproduce approximately matrix $A$, as illustrated in figure 2.

Latent Dirichlet Allocation (LDA) learns the relationships between words, topics, and documents by the assumption that documents are generated by a probabilistic model. Let $K$ be the specified number of topics, $N$ is the number of words in the corpus and two hyper parameters $\alpha$ and $\eta$ that guide the distributions. Each topic $z$ is associated with a multinomial distribution over the vocabulary $\beta_k$, drawn from $Dir(\eta)$. A given document $d_i$ in total $D$ documents, is then generated by the following process:

— Choose $\Theta_d \sim Dir(\alpha)$, a topic distribution for $d_i$
— For each word
  — Select a topic $z_{d,n} \sim \Theta_d$
  — Select the word $w_{d,n} \sim \beta_{z_{d,n}}$

In LDA model, the $\beta$ and $\Theta$ distributions represent the probability of words used in each topic and the probability of each topic appearing in each document, which correspond to word to topic matrix $W$ and document to topic matrix $H$, respectively. The process above defines a generative model corresponds to the following joint distribution, specifies a number of dependencies of the hidden and observed variables:

$$\prod_{i=1}^{K} p(\beta_i | \eta) \prod_{d=1}^{D} p(\Theta_d | \alpha) \left( \prod_{n=1}^{N} p(z_{d,n} | \Theta_d) p(w_{d,n} | \beta_{1:K}, z_{d,n}) \right) \quad (2)$$

Non-Negative Matrix Factorization (NMF) also factorizes $A$ into two lower rank matrices $W$ and $H$ with the goal to reproduce the original matrix with the lowest error, with only one constraint: the decomposed matrices consist of only non-negative values. Using the Frobenius norm for matrices (a distance measure between two given matrices), the approximation of $A$ is achieved by minimizing the error function:

$$\min_{W,H} \|A - WH\|_F, \text{ s.t } W \geq 0, H \geq 0 \quad (3)$$

NMF algorithm has fewer parameter choices involved in the modeling process than those of LDA. NMF uses an iterative process to update the initial values of $W$ and $H$. The process stops when the approximation error converges or it reaches the specified number of iterations. The algorithm randomly initializes each matrix with non-negative values, iterate for each element index $c, j$, and $i$, and modify them as the following multiplicative update rule:

$$H_{cj} \leftarrow H_{cj} \frac{(W^T A)_{cj}}{(W^T W H)_{cj}}; \; W_{ic} \leftarrow W_{ic} \frac{(A H^T)_{ic}}{(W H H^T)_{ic}} \quad (4)$$

After the matrices are trained, an unseen document is inferenced by a fold-in step, in which the document to topic matrix $H$ is partially updated while the word to topic matrix $W$ is kept, to get the distribution of topics in that document.

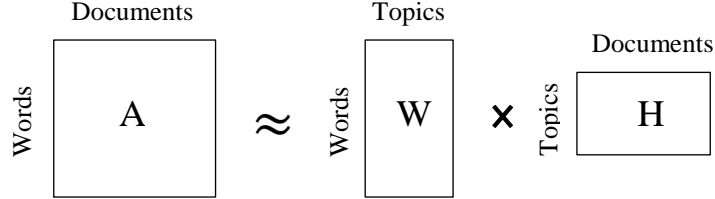

**Fig. 2.** Matrix decomposition for LDA and NMF

### 3.3 Description Similarity based Ranking

As previously described by using LDA or NMF, the bag-of-words matrix is decomposed into two lower rank matrices. The document to topic matrix $H$ is used to represent the descriptions and to compute the similarity score between two descriptions. Based on this and the category of the item, the websites are either ranked by the quantity or average price of similar items that are posted on each site.

The cosine similarity [13] is used as a distance metric to calculate the similarity score between two documents. Given two $k$ dimension vectors $d_i$ and $d_j$, which represent the distribution of $k$ topics in two different descriptions, the value of the cosine angle between them is a real number in the range [0,1]. Two vectors with the same orientation have a cosine similarity of 1 and two vectors which are perpendicular have a similarity of 0. The cosine angle between them can be calculated as follows:

$$cos(d_i, d_j) = \frac{\sum_k (d_i[k] \cdot d_j[k])}{\sqrt{\sum_k d_i[k]^2} \cdot \sqrt{\sum_k d_j[k]^2}} \quad (5)$$

For calculating the average price, after having a list of the items which have similar descriptions, we use the measures of data dispersion to get the range of price data. Let $x_1, x_2, \ldots, x_n$ be a set of ordered observations for the price attribute. The lower quartile $Q_1$, which cuts off the lowest 25% of the data and the upper quartile $Q_3$, which cuts off the lowest 75% of the data. We use the distance between the first and third quartiles which cover the middle half of the data to calculate the average price of the items whose prices fall into this range.

### 3.4 Voting Scheme based Ranking

Since users may not only want to know the selling websites by the statistics of similar descriptions but they also want other aspects such as desired selling price to be taken into account, the recommendation system also has to deal with this scenario. Based on the pre-processed features: computed semantic features, along with category feature and price feature of the training dataset, the system learns to recommend a ranked list of suitable websites for users to sell their items. This problem can be considered as classification process. However, this is a non-standard classification task. The classifier must learn how to rank a list of labels for each of the input based on a single-label training dataset. Besides, the training dataset is informal in the way that the same instance could belong to different corresponding labels, due to the fact that a user may sell his item on more than one website. We propose the ensemble learning methods [6, 7, 8], specifically Random Forests method [16], where we utilize the voting principle of the algorithm to handle the recommendation.

The basic idea of ensemble learning methods is that by combining lots of learners, some learns certain things well and some learns others so that they perform in different ways, the generated results will be significantly better than any single learner. One of the approach to combine learners is bagging [16], which stands for bootstrap aggregating. Bagging is sampling the original dataset with replacement, so that some data points in the original dataset may be replicated in the sample, whereas other points may be absent from the sample. The bootstrap sample is the same size as the original, and many of the samples are taken. This method get each learner performs slightly differently, which achieves the requirement of ensemble methods. This is also what we desire, each learner learns a different aspect and to avoid overfitting on such informal dataset.

Random Forests is a meta-learner, belongs to the ensemble learning methods, which consists of many individual decision trees [16], that are used for classification using the information entropy concept. This tree based learner requires no input preparation which is suitable to handle our mixed-type data: categorical features and numerical features in different scales. In Random Forests, there are two different sources of randomness involved in the process of building trees. First, it uses bagging, each tree is trained on different random subset of data. Second, the random selection of the attributes to split at each node, which is only the subset of the whole set of features. Commonly, the subset size is equal to the square root of the number of all features. The randomness reduces the variance without effecting the bias. Also, it speeds up the training when at each node, there is fewer features to compute. The training algorithm is described as follows:

- For each of the $N$ trees
    - Draw a bootstrap sample from the training data to train a decision tree
    - At each node, select a subset of $m$ features
    - Pick the best split point among $m$ features and split the node into child nodes
    - Repeat until the tree in complete

The algorithm works well on large dataset since each tree is independent of each other, trees can be trained in parallel. After the trees are trained, the output of the forest is the majority vote, which resemble the governments election system, a candidate must receive a majority of electoral votes to win the election. On this basis, the websites are ranked in order of decreasing number of votes.

## 4 Experiments and Results

### 4.1 The Dataset and Evaluation Method

We used the data from Vietnam C2C e-commerce, including 7 websites published by these companies: *Cho Tot* [20], *Nhat Tao* [21], *Vat Gia* [22], *Kypernet Viet Nam* [23], *Truyen Thong So* [24], *Viet Giang* [25] and *Mua Ban* [26]. The data collected in the first three websites have items of various categories, whereas the last four websites specialize in cars, motorbikes, cameras and luxury phones, respectively. In this experiment, we get from database about 55,000 pre-processed posts in which the poor quality data [17] are removed. Each post is a record with attributes containing its product's information (title description, category and price). We also get other 1,000 posts to evaluate the proposed approaches. We asked real users to judge the ranking of the websites for each of the item. The first set is ranked based on the specific criteria: the quantity or the average price of similar items, as the test set for description similarity based ranking. The second set is ranked based on the whole input information: description, category and desired selling price, to use as the test set for voting scheme ranking.

In this paper, we used evaluation metric which is specifically designed to measure the performance of a ranking system, the Normalized Distance-based Performance Measure (NDPM) [18]. Let $C^-$ be the number of contradictory preference relations which happen when the system's ranking is opposite to the reference user's ranking for a pair, $C^u$ be number of compatible preference relations which happen when the system preferred an item to another item but user sees them equal, and $C^i$ be the total number of preferred relations, in same order, of the pairs between the system's and user's ranking. The NDPM is then given by:

$$NDPM = \frac{C^- + 0.5C^u}{C^i} \quad (6)$$

The NDPM measure gives a perfect score of 0 to systems where there is total agreement between system's ranking and user's ranking. Contradicting every reference preference relation gives the worst score of 1. Contradicting a reference preference is penalized twice as much as not predicting it.

### 4.2 Results and Interpretations

We used the NDPM metric to evaluate the performance of both the description similarity based ranking and voting scheme based ranking. The goal here is to minimize the average NDPM score on the test set.

In figure 3, we measured the performance of two topic modeling techniques, NMF and LDA, and simple BOW model on the recommendation task in which users want to sort the websites in order of a specific criterion, which is the quantity or the average price of similar items in a category. The experiment is conducted with different number of topics $n_{topic} \in \{50, 100, 150, 200, 250, 300, 350\}$ and we found that when the number of topics is increased to 250, the NDPM score for NMF method obtains the lowest score compared to both LDA and BOW in both criteria. It reaches approximately 0.16 in the quantity ranking and 0.18 in the average price ranking. On the other hand, LDA method performed worse on tests than both BOW and NMF model.

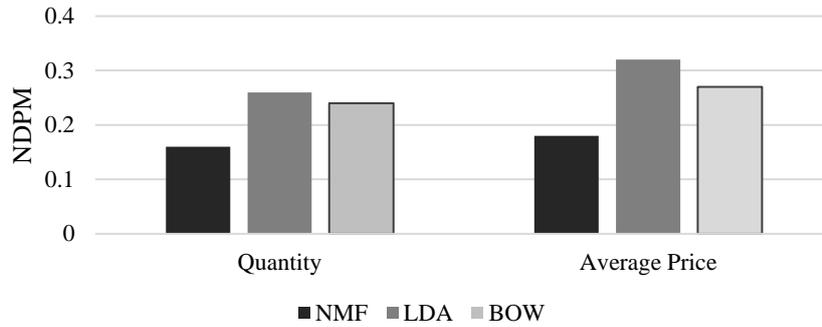

**Fig. 3.** NDPM score for three text presentation models, NMF, LDA and BOW for C2C website recommendation based on the quantity or the average price of similar items

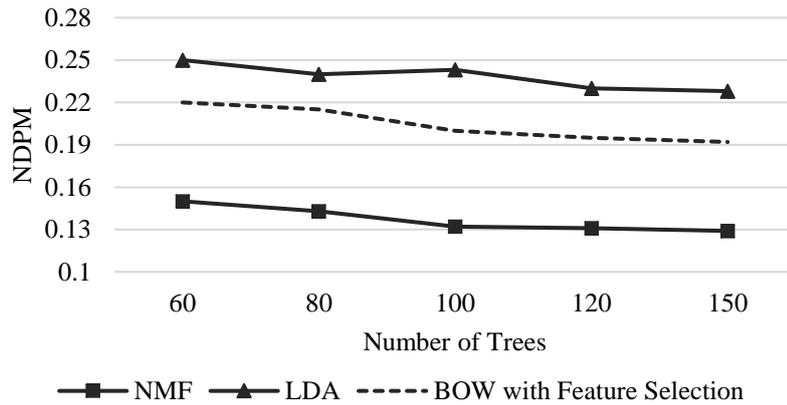

**Fig. 4.** NDPM score for C2C website recommendation based on voting rule of Random Forests

Figure 4 shows the performance of the C2C website recommendation system based on voting principle of Random Forest. The experiment is conducted on BOW method with feature selection [19] for dimensionality reduction, NMF method with 250 topics and LDA method with 100 topics, where larger numbers of topics do not significantly improve the performance of each topic model. As the graph shows, the performance of NMF method far exceeded that of LDA and BOW method. As the number of trees is increased to about 100 trees, NDPM score reaches roughly 0.13 for NMF method. Again, LDA method gave the worst performance.

As shown in table 1, we select 5 interpretable topics in 250 topics generated by NMF method as examples. We manually name each topic based on highest weighted words and highest weighted documents, which representing categories including *HTC One M smartphone, luxury phone, Sony Handycam camera, Air Blade motorbike* and *Toyota Innova car*. These topics are about categorical groups of the items of several brands without going to details about specifications of the items. Besides, there are also others topics which are not interpreted well.

**Table 1.** Top words and top documents in 5 topics discoverd by NMF method

| Topic | Top Words | Top Documents |
|---|---|---|
| Smartphone, HTC One M | one<br>htc<br>m7<br>e8<br>32gb | htc m8 gold 32gb<br>htc one m8 gold new 99%<br>htc one m7 - nguyên zin – fullbox *(htc one m7 - original – fullbox)*<br>bán htc one m7 giá rẻ *(cheap htc one m7 for sale)*<br>htc one m8 gold mới nguyên hộp *(new htc one m8 gold fullbox)* |
| Luxury phone | vertu<br>constellation<br>ti<br>ascent<br>goldvish | bán máy vertu constellation ceramic nguyên zin *(original vertu constellation ceramic for sale)*<br>vertu ascent ti chính hãng giá rẻ *(cheap and genuine vertu ascent ti)*<br>vertu constellation ayxta black alligator đẹp quý phái *(luxuriously beautiful vertu constellation ayxta black alligator)*<br>vertu constellation pink nguyên zin cần bán *(original vertu constellation pink for sale)*<br>goldvish-vertu các loại *(goldvish-vertu series)* |
| Camera, Sony Handycam | sony<br>hadycam<br>hd<br>hdr-gw77<br>cx405 | bán máy quay handycam hdr-gw77 99% *(99% handycam hdr-gw77 camera for sale)*<br>sony handycam hdr-gw77 còn mới *(like new sony handycam hdr-gw77)*<br>bán nhanh sony handycam hdr-gw77 *(sony handycam hdr-gw77 for quick sale)*<br>sony handycam cx405 xách tay fullbox *(hand-carried sony handycam cx405 fullbox)*<br>cần bán sony handycam hdr cx405, đầy đủ phụ kiện *(sony handycam hdr cx405 for sale, full of accessories)* |
| Motorbike, Air Blade | air<br>blade<br>zin<br>xe *(motorbike)*<br>2015 | air blade 125fi màu trắng xám 2015 zin 100% *(100% original grey white air blade 125fi 2015)*<br>air blade đỏ xe chính chủ máy nguyên 99% *(licensed red air blade motorbike, 99% untouched engine)* |

| Topic | Top Words | Top Documents |
|---|---|---|
| | | air blade fi đen nguyên bản 2010 *(untouched black air blade fi 2010)* |
| | | xe honda airblade chính chủ *(licensed honda air blade motorbike)* |
| | | bán airblade fi 2015 đen, máy zin giá tốt *(black air blade fi 2015 for sale, untouched engine, good price)* |
| Car, Toyota Innova | toyota innova số *(transmission)* sàn *(manual)* chủ *(licensed)* | toyota innova g đời 2010 *(toyota innova g 2010 series)* |
| | | innova g 2011 số sàn *(manual transmission innova g 2011)* |
| | | xe chính chủ toyota innova 2011 *(licensed toyota innova 2011 car)* |
| | | toyota innova 2016 xe gia đình còn zin 100% *(100% original toyota innova 2016 family car)* |
| | | xe innova 2k15 2.0G màu bạc *(silver innova 2k15 2.0G car)* |

## 5    Conclusion and Future Work

In this work, we explored the NoSQL data-based machine learning techniques in C2C e-commerce recommendation system for suggesting appropriate websites for target items, to solve the lack of similar systems. The unstructured data is first stored and indexed, then applied machine-learned models for feature representation and label ranking. We used topic modeling methods for text presentation, and utilized the voting scheme of ensemble learning methods to rank the websites. The experimental results showed that our proposed approach is effective. Using Non-Negative Matrix Factorization method for representing the distribution of topics in documents gained the best performance in both ranking based on similar descriptions and ranking based on votes of Random Forests. Specifically, the NDPM score reaches approximately 0.16, 0.18 and 0.13 in ranking according to quantity, to average price, and to whole item's information, respectively. This also showed that Random Forests, widely known for very good performance on classification task, can be used to rank labels efficiently. Further studies on applying these techniques to larger scale dataset and improving the performance by more sophisticated topic models and ranking models should be necessary.

## References

1. Konstan, J.A., Riedl, J., & Schafer, J.B.: E-commerce recommendation applications. Data Min. Knowl. Discov. **5**(1-2), 115-153 (2001)
2. Choi, I.Y., Kim, H.K., Kim, J.K., & Park, D.H.: A literature review and classification of recommender systems research. Expert Syst. Appl. **39**(11), 10059-10072 (2012)

3. Blei, D.M., Carin, L., & Dunson, D.B.: Probabilistic topic models. IEEE Signal Process. Mag. **27**(6), 55-65 (2010)
4. Arora, S., Ge, R., & Moitra, A: Learning topic models--going beyond SVD. 2012 IEEE 53rd Annual Symposium on Foundations of Computer Science, pp. 1-10 (2012)
5. Andrzejewski, D., Buttler, D., Kegelmeyer, W.P., & Stevens, K.: Exploring topic coherence over many models and many topics. 2012 Joint Conference on Empirical Methods in Natural Language Processing and Computational Natural Language Learning, pp. 952-961 (2012)
6. Dietterich, T.G.: Ensemble methods in machine learning. First International Workshop on Multiple Classifier Systems, pp. 1-15 (2000)
7. Kong, X., Shi, C., Wang, B., & Yu, P.S.: Multi-label ensemble learning. Joint European Conference on Machine Learning and Knowledge Discovery in Databases, pp. 223-239 (2011)
8. Brown, G., & Kuncheva, L. I.: "Good" and "bad" diversity in majority vote ensembles. 9th International Workshop on Multiple Classifier Systems, pp. 124-133 (2010)
9. Guangyao, C.: Research on the recommending method used in C2C online trading. 2007 IEEE/WIC/ACM International Conferences on Web Intelligence and Intelligent Agent Technology-Workshops, pp. 103-106 (2007)
10. Ai, D. X., Zuo, H., & Yang, J.: C2C e-commerce recommender system based on three-dimensional collaborative filtering. Applied Mechanics and Materials. **336**, pp. 2563-2566 (2013)
11. Bahabadi, M. D., Golpayegani, A. H., & Esmaeili, L.: A novel C2C e-commerce recommender system based on link prediction: applying social network analysis. CoRR. abs/1407.8365 (2014)
12. Kononenko, O., Baysal, O., Holmes, R., & Godfrey, M. W.: Mining modern repositories with Elasticsearch. 11th Working Conference on Mining Software Repositories, pp. 328-331 (2014)
13. Manning, C. D., Raghavan, P., & Schütze, H.: Introduction to information retrieval. Cambridge: Cambridge university press (2008)
14. Blei, D. M., Ng, A. Y., & Jordan, M. I.: Latent Dirichlet allocation. Journal of machine Learning research, **3**(Jan), 993-102 (2003)
15. Lee, D. D., & Seung, H. S.: Learning the parts of objects by non-negative matrix factorization. Nature. **401**(6755), 788-791 (1999)
16. Breiman, L.: Random Forests. Machine Learning. **45**(1), 5-32 (2001)
17. Dang, T. K., Ho, D. D., Pham, D. M. C., Vo, A. K., & Nguyen, H. H.: A cross-checking based method for fraudulent detection on e-commercial crawling data. 2016 International Conference on Advanced Computing and Applications, pp. 32-39 (2016)
18. Gunawardana, A., & Shani, G.: Evaluating recommender systems. In Recommender Systems Handbook, Springer, pp. 265-308 (2015)
19. Yang, Y., & Pedersen, J. O.: A comparative study on feature selection in text categorization. 14th International Conference on Machine Learning, pp. 412-420 (1997)
20. Cho Tot Co., Ltd: https://www.chotot.com
21. Nhat Tao E-Commerce JSC: https://nhattao.com
22. Viet Nam Price JSC: http://www.vatgia.com
23. Kypernet Viet Nam JSC: https://bonbanh.com
24. Truyen Thong So Co., Ltd: http://www.2banh.vn
25. Viet Giang Co., Ltd: http://mayanhcu.vn
26. Mua Ban JSC: https://muaban.net